# Computer Aided Optimization of the Unconventional Processing

**Assist.Prof. Tiberiu Marius Karnyanszky, Ph.D., Dipl.Eng., Dipl.Ec.**
"Tibiscus" University of Timisoara, Romania
**Assoc. Prof. MIHAI ŢÎŢU, Ph.D., Dipl. Eng., Dipl. Ec.**
„Lucian Blaga" University of Hermannstadt (Sibiu), Romania

**REZUMAT:** Tehnologiile neconvenţionale, aplicate în mod curent la o categorie vastă de materiale, dificile de prelucrat prin tehnici uzuale, au conoscut în ultimii 60 de ani toate etapele, de la descoperire şi până la utilizarea pe scară largă. Ele se bazează pe mecanismele elementare ce conduc prelucrarea prin metode clasice, dar utilizează, în plus, interconexiunile acestor metode, ceea ce duce la un câştig de performanţă prin creşterea preciziei rezultatelor, reducerea timpului de prelucrare, creşterea calităţii produsului finit etc Utilizarea calculatorului şi a unui produs software care să asiste operatorul uman în prelucrarea folosind una din metodele neconvenţionale cum sunt eroziunea electrică sau electrochimică, eroziunea complexă electrică-electrochimică, prelucrarea cu fascicul laser ş.a.m.d. nu poate decât să îmbunătăţească şi mai mult acest câştig de performanţă. Lucrarea prezintă o astfel de aplicaţie, bazată pe o bază de date conţinând rezultatele experimentale anterioare, care propune o metodă de optimizare a rezultatelor de ieşire.

## 1. Introduction

The electric erosion processing is a part of the most general category of unconventional methods, based on concentrated energies processing. The action of dimensional processing is the result of certain basic erosion processes, concentrated between the transfer object (TO) and the processed object (PO), using a hydrocarbon of the type gas-oil as dielectric, which ensure a non-conducting liquid environment.

The technological parameters are, as usually, the objective functions and can be improved by of an uniform actuation of the erosive space using a system of homogeneous, directed exterior magnetic fields, so as the quantity of removed material from PO, in the time unit, to be uniformly extracted from the crater, regarding the quantity in relation to depth and surface ([N+03]).

The study presented above contains a universal and interactive informational system which prepares the processing thorough unconventional methods, selects the experimental data, models and simulates the developing processes, optimizes and permits the running of processing process ([Kar04, N+03]), particularly applied to the electric erosion processing. The implemented data system offers a documented support for making the decisions regarding the study of the technologic process, the analysis and interpretation of the experimental data and their checking and modeling.

For the elaboration of this program, similar concepts were used in the specialty literature regarding the running of the electric erosion process ([O+99, Tit98, Nan03]), the electro-chemic erosion ([IT98, TIS99, Nan03]), the complex electric and electrochemical erosion ([Her04, Nan03]).

## 2. Experimental researches

The program is structured as following (Fig. 1):

- **Initialization** menu. It allows the erasing of all information in the database, respectively form the tables (files) PO, POPROPERTIES, TO, TOPROPERTIES, OUTCOME, INPUTS, OUTPUTS, WE, MACHINE, CLASSIC. Database structure is comprehensive presented in ([Kar06]);
- **Modification** menu. It allows the adding, the modification or the erasing of the data in the above tables;
- **Planning** menu. It allows the creating of the program matrix of a factorial experiment used to obtain input data;
- **Processing** menu. It allows the introduction, modification or erasing of the data regarding experiments, their statistic analysis, the modeling and simulation;
- **Optimizing** menu. It allows the determining the optimal conditions for the developing of a processing operation;
- **Listing** menu. It allows the obtaining of lists with information from the database;
- **Ending** menu. It allows the work with the data system ending.

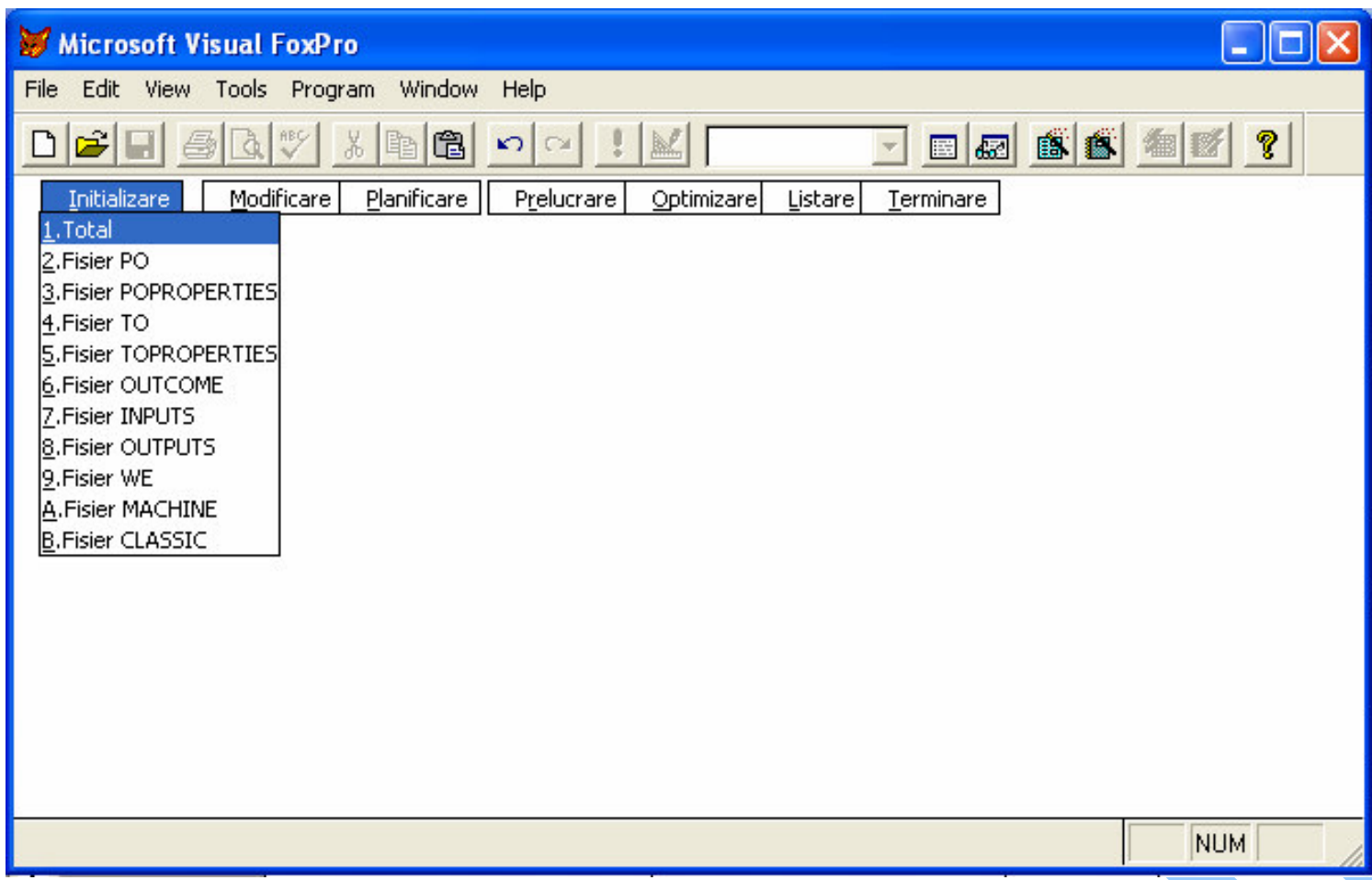

***Fig. 1:** The main menu of the application*

The introduction, the modification, the processing and the interpretation of the experimental data develop through the **Processing** menu (Fig. 2), used to:

- **Selecting data** option. It allows the introduction of experimental information in the database;
- **Statistic processing** option. It allows the analyses of data resulting from an experiment, in order to establish whether the technological parameters are homogenous or not; if they are, the processing ends; if not, the data which must be eliminated are suggested to the users and the analyzing process starts again;
- **Mono-factorial dispersion analysis** option. It allows to verify if the experimental results depend or not on one input factor;
- **Bi-factorial dispersion analysis** option. It allows to verify if the experimental results depend or not on two input factors;
- **Mono-variable modeling** option. It allows to elaborate the mathematic model of the dependence between the technologic parameter experimentally obtained (present in the database) and one input factor (also present in the database);
- **Multi-variable modeling** option. It allows to elaborate the mathematic model of the dependence between the technologic parameter experimentally obtained (present in the database) and the input factors (also from the database);

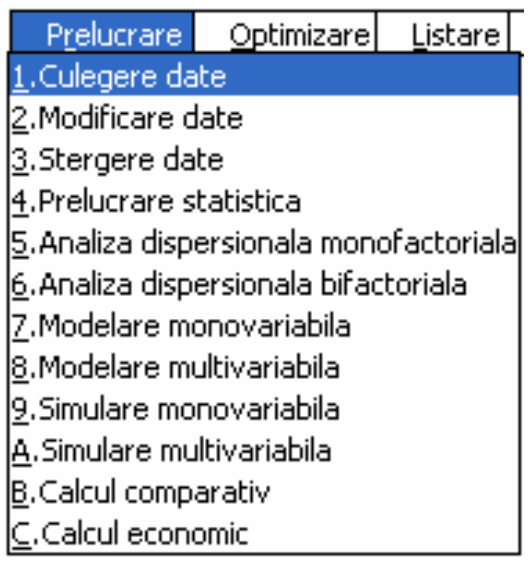

***Fig. 2:*** *Processing menu*

- **Mono-variable simulation** option. It allows the elaboration of more mathematic models and the choosing of the best of them from a large class of functions; it permits to the user to determine the performance of the processing machine in the given experimental conditions through the determination of the most suitable interpolation function of one input variable;
- **Multi-variable simulation** option. It allows the elaboration of more mathematic models and the choosing of the best of them from a large class of functions; it permits to the user to determine the performance of the processing machine in the given experimental conditions through the determination of the most suitable interpolation function of two input variables;
- **Comparative determination** option. It allows a comparison between the results of the processing through the selected unconventional method and through similar conventional methods, by re-finding of the two processing times in the data base and their comparative presentation;
- **Economic calculus** option. It allows a cost calculation of the processing through the unconventional method.

The main objective of this application is the determination of the optimal processing conditions for the materials which can be processed by a specific unconventional method, particularly the electric erosion processing method. To this purpose, the **Optimizing** menu permits the choosing of the work conditions and the introduction of the other necessary information and, then, the displaying of the results of the optimizing process (Fig. 3). The result is the best determined polynomial function that represents the dependence between one technological parameter (the evolution of the volume wear, in this example) and a few input parameters (the width inter-power impulse, the power impulse, the current and the magnetic field strength, in this example).

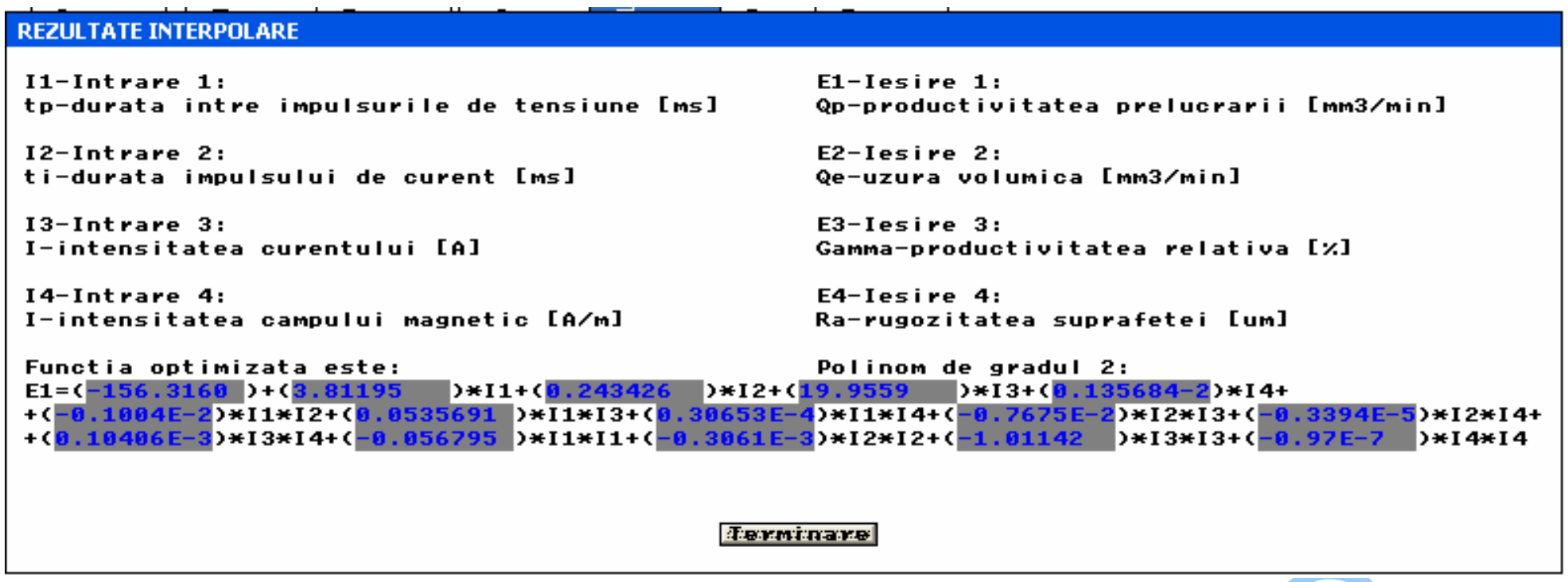

***Fig. 3:*** *The results of the optimization*

## Conclusions

There are a lot of unconventional processing methods and the electric erosion is the most developed, because all internal processes are well known, the technological systems are wide used and the practical applications are unlimited.

This is the main reason the authors elaborated the presented data system which contains the data base of the experimental results and permits to obtain information on the processing, according to a classification criterion such as: the processing machine, the material, the shape and the dimensions of the processed and transfer object, the material etc.

This system contains also the automatic establishing algorithms of the mathematic models and the optimization algorithms for each experience set present in the database. Based on this system, for a user who wants to use this technologic procedure, it is extremely helpful because it permits to choose the work conditions which supposes a minimum processing time, an optimal quality of the resulting surface, the lowest possible forces and middles expenditure.

## References

[Her04] R. I. E. Herman (coordonator) - *Tratat de tehnologii neconvenționale. Vol. V – Prelucrarea prin eroziune complexă electrică-electrochimică*, Asociația Română pentru Tehnologii Neconvenționale, Academia Română, Academia de Ştiințe Tehnice din România, Editura Augusta, Timişoara, 2004

[IT98] M. Ivan, R. Temur – *Proiectarea tehnologiilor de prelucrare prin eroziune electrochimică a suprafeţelor frontale folosind baze de date computerizate,* Revista de tehnologii neconvenţionale nr. 1/1998, Editura Augusta, Timişoara, 1998

[Kar04] T. M. Karnyanszky – *Contribuţii la conducerea automată a prelucrării dimensionale prin eroziune electrică complexă*, Teză de doctorat, Universitatea "Politehnica" din Timişoara, 2004

[Kar06] T. M. Karnyanszky – *Database Model for the Optimisation of the Processing by Unconventional Methods*, Buletinul Institutului Politehnic Iaşi, Tomul LII (LVI), fasc. 5A, 2006

[N+03] D. Nanu, M. Ţîţu, N. Pricpo, E. Hubert – *Experimental Studies and Researches on Reducing the Wera in the Electric Erosion Processing*, Revista de Tehnologii Neconvenţionale nr. 1/2003, Editura Augusta, Timişoara, 2003

[Nan03] A. Nanu (coordonator) - *Tratat de tehnologii neconvenţionale. Vol. I – Tehnologiile neconvenţionale la început de mileniu*", Asociaţia Română pentru Tehnologii Neconvenţionale, Academia Română, Academia de Ştiinţe Tehnice din România, Editura Augusta, Timişoara, 2003

[O+99] C. Oprean, D. Nanu, M. Ţîţu, E. Cicală, A. B. Vannes – *Software universal pentru modelare, optimizare şi conducere asistată a proceselor tehnologice,* Revista de tehnologii neconvenţionale nr. 1/1999, Editura Augusta, Timişoara, 1999

[TIS99] R. Temur, M. Ivan, F. Sârbu – *Utilizarea microcalculatorului pentru controlul procesului de prelucrare prin eroziune electrochimică,* Tehnologii neconvenţionale nr. 2/1999, Editura Augusta, Timişoara 1999

[Tit98] M. Ţîţu – *Contribuţii cu privire la modificarea transferului substanţial la prelucrarea dimensională prin eroziune electrică cu câmpuri coercitive*, Teză de doctorat, Universitatea "Lucian Blaga" Sibiu, 1998